\documentclass{revtex4}

\input epsf
\usepackage{amsmath,amssymb}
\usepackage{graphicx}

\draft

\begin{document}
\titlepage
\title{Inflationary attractor in Gauss-Bonnet brane cosmology}
\author{Xin-He Meng$^{1,2,3}$ \footnote{xhmeng@phys.nankai.edu.cn}
 \ \ Peng Wang$^1$ \footnote{pwang234@sohu.com}
} \affiliation{1.  Department of Physics, Nankai University,
Tianjin 300071, P.R.China \\2. Institute of Theoretical Physics,
CAS, Beijing 100080, P.R.China \\3. Department of Physics,
University of Arizona, Tucson, AZ 85721}

\begin{abstract}
The inflationary attractor properties of the canonical scalar
field and Born-Infeld field are investigated in the
Randall-Sundrum II scenario with a Gauss-Bonnet term in the bulk
action. We find that the inflationary attractor property will
always hold for both the canonical and Born-Infeld fields for any
allowed non-negative Gauss-Bonnet coupling. We also briefly
discuss the possibility of explaining the suppressed lower
multiples and running scalar spectral index simultaneously in the
scenario of Gauss-Bonnet brane inflation.
\end{abstract}

\maketitle

\textbf{1. Introduction}

Recently, there is considerable interest in inflationary models
motivated by string/M-theory (See \cite{Lidsey2} for a review and
reference therein). In particular, much attention has been focused
on the braneworld scenario, where our observable four-dimensional
universe is modelled as a domain wall embedded in a
higher-dimensional bulk space \cite{Arkani-Hamed}. An important
realization of this picture is provided by the Randall-Sundrum
type II scenario (RSII), where a spatially isotropic and
homogenous brane propagates in a 5-dimensional bulk space with an
effectively negative cosmological constant \cite{RSII}. This
scenario has many interesting cosmological implications. The
standard Friedmann equation would be modified when the energy
density of the matter confined to the brane is larger than the
brane tension \cite{RSMF}. This Modified Friedmann equation will
lead to a faster Hubble expansion and a more strongly damped
evolution of the scalar field. This assists the slow-roll
inflation, enhances the amount of inflation obtained in any given
inflation model, and drives the perturbations towards an exactly
scale-invariant Harrison-Zel'dovich spectrum. Inflation in RSII
model has been studied for both canonical scalar field (see e.g.
\cite{Maartens}) and Born-Infeld field (see e.g. \cite{Bento}).

One approach to developing the braneworld scenario in a more
string theoretic setting is to introduce higher-order curvature
invariants in the bulk action \cite{Odintsov-hd,Char,Davis,Gra}.
Specifically, the Gauss-Bonnet combination arises as the leading
order for quantum corrections in the heterotic string effective
action and in five dimensions it represents the unique combination
of curvature invariants that leads to second-order field equations
in the metric tensor \cite{GB}. Note that quantum effects on the
brane also induce higher curvature terms \cite{Odintsov-gb},
where, however, the Gauss-Bonnet term is topological invariant. In
the RSII scenario with a Gauss-Bonnet term, the effective
Friedmann equation describing the motion of brane containing
general perfect fluid may be derived from a generalization of the
Birkhoff's theorem \cite{Char}. A more geometrical approach may be
taken by varying the boundary term \cite{Davis} which was
calculated by Myers \cite{Myers} or employing the formalism of
differential forms \cite{Gra}. Braneworld holography in
Gauss-Bonnet gravity was studied in Ref.\cite{gregory}, where it
was demonstrated that a holographic description for braneworld
matter could be found in the Gauss-Bonnet bulk.

Thus, the investigation of the effects of a Gauss-Bonnet term on
inflationary braneworld models in the RSII scenario is well
motivated. This was analyzed in detail first in Ref.\cite{Lidsey},
where the authors have derived the expressions for slow-roll
parameters and spectral indices, etc. However, up to now, some
features of Gauss-Bonnet brane inflation have not been
investigated in detail in the literature (See Ref.\cite{sami} for
a recent discussion). Specifically, if inflation is to be truly
predictive, the evolution when the scalar field is at some given
point on the potential has to be independent of the initial
conditions. Otherwise, any result, such as the amplitude of
density perturbations, would depend on the unknown initial
conditions. However, the equation of motion for the scalar field
is  second order, implying that $\dot \phi$ can in principle take
on any value anywhere on the potential provided, and so there
certainly is not a unique solution at each point on the potential.
Inflation can therefore only be predictive if the solution
exhibits an attractor behavior, where the differences between
solutions of different initial conditions rapidly vanish
\cite{Brandenberger} (See Sec.3.7 of Ref.\cite{Liddle} for a
review). Thus it is important to investigate whether the
inflationary attractor property holds for Gauss-Bonnet brane
inflation driven by the canonical scalar field or Born-Infeld
field. In RSII scenario without the Gauss-Bonnet contribution, the
inflationary attractor property is shown to be held for both the
canonical scalar field and Born-Infeld field \cite{Zhang1}. In the
standard FRW cosmology, the inflationary attractor property is
shown to be held for Born-Infeld field in Ref.\cite{Piao}. In this
paper, we will show that the attractor property still holds for
both the canonical and Born-Infeld fields in the Gauss-Bonnet
braneworld scenario. The conclusions we get in this paper is quite
general and model-independent.

\textbf{2. Modified Friedmann equation in Gauss-Bonnet brane
cosmology}

In this section, we will follow closely Ref.\cite{Lidsey}. The
5-dimensional bulk action for the Gauss-Bonnet braneworld scenario
is given by
\begin{equation}
S=\frac{1}{2\kappa_5^2}\int_\mathcal{M}
d^5x\sqrt{-g}[R-2\Lambda+\alpha(R^2-4R_{AB}R^{AB}+R_{ABCD}R^{ABCD})]+S_{\partial
\mathcal{M}}+S_m\label{2.1}
\end{equation}
where $\alpha\ge0$ with dimension $(mass)^2$ is the Gauss-Bonnet
coupling, $\Lambda<0$ is the bulk cosmological constant and
$\kappa_5^2=8\pi M_5^{-3}$ determines the 5d Planck scale. The
$S_{\partial \mathcal{M}}$ is the boundary term that is required
to cancel normal derivatives of the metric tensor which arises
when varying the action with respect to the metric \cite{Myers},
and the $S_m$ is the action for matter confined on the brane. We
assume a $\mathbf{Z}_2$ symmetry across the bane. We will consider
the case that a perfect fluid matter source with density $\rho$ is
confined to the brane and the brane have a tension $\lambda$.

Define a new constant, $b$:
\begin{equation}
b\equiv(1+\frac{4}{3}\alpha\Lambda)^{3/2}\label{2.2}
\end{equation}
Note that from Eq.(\ref{2.2}), since $\Lambda<0$, there is an
upper bound on $\alpha$
\begin{equation}
\alpha\leq\frac{3}{4|\Lambda|}\equiv \alpha_u\label{2.61}
\end{equation}

In order that the standard Friedmann equation be recovered at
sufficiently low energy scales $(\rho\ll \lambda)$, we have the
identification
\begin{equation}
\kappa_4^2\equiv\frac{8\pi}{M_4^2}=\frac{\kappa_5^4\lambda}{6b^{2/3}}\label{2.3}
\end{equation}
where $M_4$ is the 4d Planck scale.

Introducing a dimensionless variable $x$, the Modified Friedmann
equations are
\begin{equation}
\rho+\lambda=(\frac{\lambda
b^{1/3}}{3\alpha\kappa_4^2})^{1/2}\sinh x\label{2.62}
\end{equation}
\begin{equation}
H^2=\frac{1}{4\alpha}[b^{1/3}\cosh (\frac{2x}{3})-1]\label{2.6}
\end{equation}
We assume that during inflation, $\rho\gg\lambda$, since this is
the case that the Modified Friedmann equation is different from
the standard one. Thus we can replace Eq.(\ref{2.62}) by

\begin{equation}
\rho=(\frac{\lambda b^{1/3}}{3\alpha\kappa_4^2})^{1/2}\sinh
x\label{2.5}
\end{equation}
Note $\rho$ increases monotonically with $x$.

In order for that the 4d effective cosmological constant vanishes,
the brane tension should satisfy
\begin{equation}
\lambda=\frac{3}{2}\frac{1-b^{1/3}}{\alpha\kappa_4^2}\label{2.7}
\end{equation}
It is easy to see that $\lambda\sim |\Lambda|/\kappa_4^2$ for any
$0\leq\alpha\leq\alpha_u$.

\textbf{3. Attractor property for canonical scalar field}

We consider a canonical scalar field confined on the brane. It is
described by the Lagrangian
\begin{equation}
L_{canonical}=-\frac{1}{2}(\partial_\mu\phi)^2-V(\phi)\label{3.1}
\end{equation}
where we use the metric signature $\{-,+,+,+\}$. In a spatially
flat FRW universe, we can assume that $\phi$ is spatially
homogeneous. The energy density and the pressure are given by
\begin{equation}
\rho_\phi=\frac{1}{2}\dot\phi^2+V(\phi)\label{3.2}
\end{equation}
\begin{equation}
p_\phi=\frac{1}{2}\dot\phi^2-V(\phi)\label{3.3}
\end{equation}

The evolution equation of the field $\phi$ is
\begin{equation}
\ddot\phi+3H\dot\phi+V'(\phi)=0\label{3.4}
\end{equation}
where a prime denotes differentiation with respect to $\phi$.

In the high energy limit, i.e. $\rho_\phi\gg\lambda$,
Eq.(\ref{2.5}) can be written as
\begin{equation}
\frac{1}{2}\dot\phi^2+V(\phi)=(\frac{\lambda
b^{1/3}}{3\alpha\kappa_4^2})^{1/2}\sinh x\label{3.5}
\end{equation}

In analyzing the inflationary attractor property, we will use the
Hamilton-Jacobi formulation of the Friedmann equation \cite{HJ}.
In this formulation, we will view the scalar field $\phi$ as the
time variable. This requires that the $\phi$ field does not change
sign during inflation. Without loss of generality, we can choose
$\dot\phi>0$ in the following discussions.

Differentiating Eq.(\ref{2.6}) with respect to $t$ gives
\begin{equation}
H\dot H=\frac{b^{1/3}}{12\alpha}\sinh (\frac{2x}{3})\dot
x\label{3.51}
\end{equation}

Differentiating Eq.(\ref{3.5}) with respect to $t$ and using
Eq.(\ref{3.4}) give
\begin{equation}
\dot x=-3H(\frac{3\alpha\kappa_4^2}{\lambda
b^{1/3}})^{1/2}\frac{\dot\phi^2}{\cosh x}\label{3.6}
\end{equation}

Substituting Eq.(\ref{3.6}) into Eq.(\ref{3.51}), we have
\begin{equation}
H'(\phi)=-\frac{b^{1/3}}{4\alpha}(\frac{3\alpha\kappa_4^2}{\lambda
b^{1/3}})^{1/2}\frac{\sinh \frac{2x}{3}}{\cosh
x}\dot\phi\label{3.7}
\end{equation}

Substituting this equation into Eq.(\ref{3.5}) gives
\begin{equation}
\frac{8\alpha\lambda}{3b^{1/3}\kappa_4^2}(\frac{\cosh x}{\sinh
\frac{2x}{3}})^2H'(\phi)^2+V(\phi)=(\frac{\lambda
b^{1/3}}{3\alpha\kappa_4^2})^{1/2}\sinh x\label{3.8}
\end{equation}

Eqs.(\ref{3.7}), (\ref{3.8}) and (\ref{2.6}) are the
Hamilton-Jacobi formulation of the Modified Friedmann equations,
which is more convenient to be used in analyzing the inflationary
attractor behaviors than Eqs.(\ref{3.4}), (\ref{2.5}) and
(\ref{2.6}). In this formulation, one considers $H(\phi)$, rather
than $V(\phi)$, as the fundamental quantities. If we can solve
$H(\phi)$ from Eqs.(\ref{3.7}) and (\ref{2.6}), substituting into
Eqs.(\ref{3.8}) and (\ref{2.6}) we can immediately obtain
$V(\phi)$. Therefore, the Hamilton-Jacobi formalism is also very
useful to obtain a large set of exact inflationary solution (See
Ref.\cite{Zhang1} for some examples) and put general constraints
on the form of the potentials \cite{Liddle2}.

Supposing $(H_0(\phi), x_0)$ is any solution to Eqs.(\ref{3.8})
and (\ref{2.6}), which can be either inflationary or
non-inflationary. We will consider the homogeneous perturbations
$(\delta H(\phi), \delta x)$ to this solution. The attractor
property will be satisfied if it becomes small as $\phi$
increases. Substituting $H(\phi)=H_0(\phi)+\delta H(\phi)$ and
$x=x_0+\delta x$ into Eqs.(\ref{3.8}), (\ref{2.6}) and
linearizing, we find that the perturbations obey
\begin{equation}
\delta x=\frac{12\alpha H_0(\phi)}{b^{1/3}\sinh
\frac{2x_0}{3}}\delta H(\phi)\label{3.9}
\end{equation}
\begin{equation}
 H_0'(\phi)\delta H'(\phi)=g(\phi)\delta
x\label{3.10}
\end{equation}
where $g(\phi)$ is given by
\begin{equation}
g(\phi)=\frac{3b^{1/3}\kappa_4^2}{16\alpha\lambda}(\frac{\lambda
b^{1/3}}{3\alpha\kappa_4^2})^{1/2}\frac{(\sinh
\frac{2x_0}{3})^2}{\cosh x_0}-(\tanh x_0-\frac{2}{3}\coth
\frac{2x_0}{3})H_0'(\phi)^2\label{3.102}
\end{equation}

Eqs.(\ref{3.9}), (\ref{3.10}) can be integrated to give
\begin{equation}
\delta H(\phi)=\delta H(\phi_i)\exp[\frac{12\alpha}{b^{1/3}\sinh
\frac{2x_0}{3}}\int^\phi_{\phi_i}g(\phi)\frac{H_0(\phi)}{
H_0'(\phi)}d\phi]\label{3.101}
\end{equation}
where $\delta H(\phi_i)$ is the value at some initial point
$\phi_i$.

Since $H_0'$ and $d\phi$ have the opposite sign, if $H_0$ is an
inflationary solution, the perturbations will damp exponentially
if $g(\phi)>0$, i.e.
\begin{equation}
\frac{3b^{1/3}\kappa_4^2}{16\alpha\lambda}(\frac{\lambda
b^{1/3}}{3\alpha\kappa_4^2})^{1/2}\frac{(\sinh
\frac{2x_0}{3})^2}{\cosh x_0}-(\tanh x_0-\frac{2}{3}\coth
\frac{2x_0}{3})H_0'(\phi)^2>0\label{3.11}
\end{equation}

Using Eqs.(\ref{3.4}) and (\ref{3.7}), this can be written as
\begin{equation}
\coth x_0 V(\phi)>(\tanh
x_0-\frac{2}{3}\coth\frac{2x_0}{3}-\frac{1}{2}\coth
x_0)\dot\phi^2\label{3.12}
\end{equation}

Since $\tanh x_0-\frac{2}{3}\coth\frac{2x_0}{3}-\frac{1}{2}\coth
x_0<0$ for all $x_0>0$, this inequality is always satisfied. Thus
we conclude that the inflationary attractor property will hold for
canonical scalar field for any $0\leq\alpha\leq\alpha_u$.

\textbf{Remark:} If considering the phantom inflation \cite{piao,
Wang}, it has been shown in Ref.\cite{Wang} that in the RSII
scenario the inflationary attractor property does not hold. It can
be shown by almost the same derivation of this section that the
inflationary attractor still does not hold in the Gauss-Bonnet
braneworld for any $\alpha$. Thus, from the point of view of the
inflationary attractor property, the scenario of phantom inflation
is unappealing.

\textbf{4. Attractor property for Born-Infeld  field}

The effective Lagrangian for tachyon is the Born-Infeld action
\cite{garo,Sen}
\begin{equation}
L_{BI}=-V(\phi)\sqrt{1+(\partial_\mu\phi)^2}\label{4.1}
\end{equation}

In a spatially flat FRW universe, we can assume that $\phi$ is
spatially homogeneous. The energy density and the pressure are
given by
\begin{equation}
\rho_\phi=\frac{V(\phi)}{\sqrt{1-\dot\phi^2}}\label{4.2}
\end{equation}
\begin{equation}
p_\phi=-V(\phi)\sqrt{1-\dot\phi^2}\label{4.3}
\end{equation}

The evolution equation of $\phi$ is
\begin{equation}
\frac{\ddot\phi}{1-\dot\phi^2}+3H\dot\phi+\frac{V'(\phi)}{V(\phi)}=0\label{4.4}
\end{equation}

In the high energy limit, i.e. $\rho_\phi\gg\lambda$,
Eq.(\ref{2.5}) can be written as
\begin{equation}
\frac{V(\phi)}{\sqrt{1-\dot\phi^2}}=(\frac{\lambda
b^{1/3}}{3\alpha\kappa_4^2})^{1/2}\sinh x\label{4.5}
\end{equation}

Differentiating Eq.(\ref{4.5}) with respect to $t$ and using
Eq.(\ref{4.4}) give
\begin{equation}
\dot x=-3H(\frac{3\alpha\kappa_4^2}{\lambda
b^{1/3}})^{1/2}\frac{V(\phi)\dot\phi^2}{\cosh
x\sqrt{1-\dot\phi^2}}\label{4.6}
\end{equation}

Substituting Eq.(\ref{4.6}) into Eq.(\ref{3.51}) gives
\begin{equation}
H'(\phi)=-\frac{b^{1/3}}{4\alpha}(\frac{3\alpha\kappa_4^2}{\lambda
b^{1/3}})^{1/2}\frac{\sinh \frac{2x}{3}}{\cosh
x}\frac{V(\phi)\dot\phi}{\sqrt{1-\dot\phi^2}}\label{4.7}
\end{equation}

Substituting this equation into Eq.(\ref{4.5}), we can get
\begin{equation}
\frac{16\alpha\lambda}{3b^{1/3}\kappa_4^2}(\frac{\cosh x}{\sinh
\frac{2x}{3}})^2H'(\phi)^2+V(\phi)^2=\frac{\lambda
b^{1/3}}{3\alpha\kappa_4^2}\sinh^2 x\label{4.8}
\end{equation}

Eqs.(\ref{4.7}), (\ref{4.8}) and (\ref{2.6}) are the
Hamilton-Jacobi formulation of the Modified Friedmann equations.

As in the previous section, assuming $(H_0(\phi), x_0)$ is any
solution to Eqs.(\ref{4.8}) and (\ref{2.6}), substituting
$H(\phi)=H_0(\phi)+\delta H(\phi)$ and $x=x_0+\delta x$ into
Eqs.(\ref{4.8}), (\ref{2.6}) and linearizing, we find that the
perturbations obey
\begin{equation}
\delta x=\frac{12\alpha H_0(\phi)}{b^{1/3}\sinh
\frac{2x_0}{3}}\delta H(\phi)\label{4.9}
\end{equation}
\begin{equation}
H_0'(\phi)\delta H'(\phi)=h(\phi)\delta x\label{4.10}
\end{equation}
where $h(\phi)$ is given by
\begin{equation}
h(\phi)=\frac{b^{2/3}}{16\alpha^2}\frac{\sinh x_0\sinh^2
\frac{2x_0}{3}}{\cosh x_0}-(\tanh x_0-\frac{2}{3}\coth
\frac{2x_0}{3})H_0'(\phi)^2\label{4.101}
\end{equation}

Eqs.(\ref{4.9}), (\ref{4.10}) can be integrated to give
\begin{equation}
\delta H(\phi)=\delta H(\phi_i)\exp[\frac{12\alpha}{b^{1/3}\sinh
\frac{2x_0}{3}}\int^\phi_{\phi_i}h(\phi)\frac{H_0(\phi)}{
H_0'(\phi)}d\phi]\label{3.101}
\end{equation}
where $\delta H(\phi_i)$ is the value at some initial point
$\phi_i$.

Since $H_0'$ and $d\phi$ have the opposite sign, if $H_0$ is an
inflationary solution, the perturbations will damp exponentially
if $h(\phi)>0$, i.e.
\begin{equation}
\frac{b^{2/3}}{16\alpha^2}\frac{\sinh x_0\sinh^2
\frac{2x_0}{3}}{\cosh x_0}-(\tanh x_0-\frac{2}{3}\coth
\frac{2x_0}{3})H_0'(\phi)^2>0\label{4.11}
\end{equation}

Using Eqs.(\ref{4.5}), (\ref{4.7}), this can be written as
\begin{equation}
\coth x_0>(\tanh x_0-\frac{2}{3}\coth
\frac{2x_0}{3})\dot\phi^2\label{4.12}
\end{equation}

Note from the action (\ref{4.1}), $\dot\phi^2\leq1$. In presence
of this constraint on $\dot\phi^2$, it is easy to check that
inequality (\ref{4.12}) is always satisfied. Thus the inflationary
attractor property will hold for Born-Infeld tachyon for any
$0\leq\alpha\leq\alpha_{u}$.

\textbf{Remark}: If considering the Born-Infeld phantom field
\cite{Li1}, it has been shown in Ref.\cite{Wang} that in the RSII
scenario the inflationary attractor property does not hold. It can
be shown by almost the same derivation of this section that the
condition for the inflationary attractor property to hold for
Born-Infeld phantom is $\alpha_c\le\alpha\le\alpha_u$ where
$\alpha_{c}\sim \frac{\lambda}{\kappa_4^2\rho_\phi^2}$. Thus it is
interesting to see that the presence of a Gauss-Bonnet term in the
bulk action can restore the inflationary attractor property of
Born-Infeld phantom.

\textbf{5. Conclusions and discussions}

In this paper, we have derived the Hamilton-Jacobi equations of
inflation in the Randall-Sundrum II scenario with a Gauss-Bonnet
contribution in the bulk action driven by a canonical or
Born-Infeld scalar fields. Using those equations, we studied the
inflationary attractor properties of the canonical and Born-Infeld
scalar fields. We found that the inflationary attractor property
holds for both the canonical and Born-Infeld fields for all the
allowed non-negative Gauss-Bonnet coupling constant. This
justifies that it is sensible to discuss Gauss-Bonnet inflation
driven by either canonical or Born-Infeld field without a
fine-tuning of the initial conditions of the inflaton field in
order that the inflation happens.

Finally, we will present an observation of an interesting feature
of the MF equations (\ref{2.5}) and (\ref{2.6}), which makes the
scenario of inflation in Gauss-Bonnet brane cosmology rather
appealing: At sufficiently high energies, i.e. $x\gg1$, the MF
equations can be approximated by $H^2\propto\rho^{2/3}$. At low
energies, i.e. $x\ll1$, the MF equations will behave like
$H^2\propto\rho^2$. In either case, the MF equation behaves like
$H^2=A\rho^q$, while $q$ will increase when the energy decreases
\cite{Lidsey}. For this form of Friedmann equation, the amplitude
of density perturbations is given by
\begin{equation}
A_s^2=\frac{9A^3V^{3q}}{25\pi^2V'^2}\label{5.2}
\end{equation}
The scalar spectral index is given by
\begin{equation}
n_s-1=\frac{1}{AV^{q-1}}(\frac{2V''}{3V}-\frac{qV'^2}{V^2})\label{5.3}
\end{equation}
Then, at high energy region, i.e. large scales, where $q=2/3<1$,
the amplitude of density perturbations will be suppressed relative
to the standard Friedmann evolution. Furthermore, when energy
scale decreases, i.e. $q$ increases, the spectral index will
decrease, i.e. the running is negative. Thus, the two important
features of CMB observations by WMAP \cite{WMAP} may be explained
simultaneously in the framework of Gauss-Bonnet brane cosmology
without a changing of the potentials or tuning of the initial
conditions \cite{Linde}. Here, those features are due to
modification of cosmic dynamics in early universe! This is a
rather elegant possibility. In this work, we have focused on the
model-independent features of the Gauss-Bonnet brane inflation.
Next, it is interesting to see whether some specific model can
accommodate the data easily.

\textbf{Acknowledgements}

We would like to thank S.D.Odintsov and S.Nojiri for valuable
comments and suggestions on the first draft of the manuscript. We
would like to thank I.P.Neupane for valuable comments on the first
uploaded version of this paper. We would also like to thank D.Lyth
and N.J.Nunes for helpful discussions. This work is partly
supported by Doctoral Foundation of National Education Ministry
and ICSC-World Laboratory Scholarship.


\begin{thebibliography}{99}
\bibitem{Lidsey2} J.E.Lidsey, astro-ph/0305528; J.E.Lidsey,
D.Wands and E.J.Copeland, Phys.Rep. \textbf{337} (2000) 343;
M.Gasperini and G.Veneziano, Phys.Rep. \textbf{373} (2003) 1;
M.Quevedo, Class.Quant.Grav. \textbf{19} (2002) 5721
[hep-th/0210292];
\bibitem{Arkani-Hamed} N.Arkani-Hamed, S.Dimopoulos and G.Dvali,
Phys.Lett. \textbf{429} (1998) 263; I.Antoniadis, N.Arkani-Hamed,
S.Dimopoulos and G.Dvali, Phys.Lett. \textbf{436} (1998) 257;
L.Randall and R.Sundrum, Phys.Rev.Lett. \textbf{83} (1999) 3370;
\bibitem{RSII} L.Randall and R.Sundrum, Phys.Rev.Lett. \textbf{83}
(1999) 4690;
\bibitem{RSMF} P.Binetruy, C.Deffayet, U.Ellwanger and D.Langlois,
Phys.Lett. \textbf{B477} (2000) 285 [hep-ph/9910219];
\bibitem{Maartens} R.Maartens, D.Wands, B.A.Bassett and
I.P.C.Heard, Phys.Rev. \textbf{D62} (2000) 041301
[hep-ph/9912464]; E.J.Copeland, A.R.Liddle and J.E.Lidsey,
Phys.Rev. \textbf{D64} (2001) 023509;
\bibitem{Bento} M.C.Bento, O.Bertolami and A.A.Sen, Phys.Rev.
\textbf{D67} (2003) 063511 [hep-ph/0208124]; S. Mukohyama,
Phys.Rev. \textbf{D66} (2002) 024009 [hep-th/0204084];

\bibitem{Odintsov-hd} S.Nojiri and S.D.Odintsov, JHEP \textbf{0007}
(2000) 049; S.Nojiri, S.D.Odintsov and S.Ogushi, Int.J.Mod.Phys.
\textbf{A16} (2001) 5085; S.Nojiri, S.D.Odintsov and S.Ogushi,
Phys.Rev. \textbf{D65} (2002) 023521; J.E.Lidsey, S.Nojiri and
S.D.Odintsov, JHEP \textbf{06} (2002) 026; J.E.Kim, B.Kyae and
H.M.Lee, Phys.Rev. \textbf{D62} (2000) 045013; I.P.Neupane, JHEP
\textbf{0009} (2000) 040 [hep-th/0008190]; I.P.Neupane, Phys.Lett.
\textbf{B512} (2001) 137 [hep-th/0104226]; Y. M. Cho, I.P. Neupane
and P. S. Wesson, Nucl.Phys. \textbf{B621} (2002) 388
[hep-th/0104227]; I.P.Neupane, Class.Quant.Grav. \textbf{19}
(2002) 5507 [hep-th/0106100]; Y. M. Cho, I. P. Neupane,
Int.J.Mod.Phys. \textbf{A18} (2003) 2703 [hep-th/0112227];
\bibitem{Char} C.Charmousis and J.Dufaux, Class.Quant.Grav. \textbf{19} (2002)
4671 [hep-th/0202107];
\bibitem{Davis} S.C.Davis, Phys.Rev. \textbf{D67} (2003) 024030
[hep-th/0208205];
\bibitem{Myers} R.C.Myers, Phys.Rev. \textbf{D36} (1987) 392;
\bibitem{Gra} E.Gravanis and S.Willison, Phys.Lett. B562 (2003)
118 [hep-th/0209076];
\bibitem{gregory} J. P. Gregory and A. Padilla, Class.Quant.Grav. \textbf{20} (2003)
4221 [hep-th/0304250];
\bibitem{Odintsov-gb} S.Nojiri and S.D.Odintsov, Phys.Lett.
\textbf{484} (2000) 119 [hep-th/0006232];
\bibitem{GB} B.Zweibach, Phys.Lett. \textbf{B156} (1985) 315;
D.Lovelock, J.Math.Phys. \textbf{12} (1971) 498;
\bibitem{Lidsey} J.E.Lidsey and N.J.Nunes, Phys.Rev. \textbf{D67} (2003)
103510 [astro-ph/0303168];
\bibitem{sami} B.C. Paul and M. Sami, hep-th/0312081;
\bibitem{Brandenberger} D.S.Goldwirth, Phys.Lett. \textbf{B243} (1990) 41; R.Brandenberger, G.Geshnizjani and
S.Watson, Phys.Rev. \textbf{D67} (2003) 123510 [hep-th/0302222];
\bibitem{Liddle} A.R.Lidde and D.H.Lyth, Cosmological Inflation and Large
Scale Structure, Cambrigde University Press, 2000;
\bibitem{HJ} D.S.Salopek and J.R.Bond, Phys.Rev. \textbf{D42}
(1990) 3936; A.G.Muslimov, Class.Quant.Grav. \textbf{7} (1990)
231; J.E.Lidsey, Phys.Lett. \textbf{B273} (1991) 42; A.R.Liddle,
P.Parsons and J.D.Barrow, Phys.Rev. \textbf{D50} (1994) 7222
[astro-ph/9408015];
\bibitem{Zhang1} Z.K.Guo, H.S.Zhang and Y.Z.Zhang, hep-ph/0309163;
\bibitem{Liddle2} A.Vallinotto, E.J.Copeland, E.W.Kolb and
A.R.Liddle, astro-ph/0311005;
\bibitem{Piao} Z.-K. Guo, Y.-S. Piao, R.-G. Cai and Y.-Z. Zhang, Phys.Rev. \textbf{D68} (2003)
043508;

\bibitem{garo} M.R.Garousi, Nucl.Phys. \textbf{B584} (2000)
284-299 [hep-th/0003122];
\bibitem{Sen} A.Sen, JHEP \textbf{0204} (2002) 048; ibid, JHEP
\textbf{0207} (2002) 065;
\bibitem{carroll} S.M.Carroll and M.Kaplinghat, Phys.Rev. \textbf{D65} (2002) 063507 [astro-ph/0108002];
\bibitem{Li1} J.G.Hao and X.Z.Li, Phys.Rev. \textbf{D68} (2003) 043501
[hep-th/0305207]; D.J.Liu and X.Z.Li, Phys.Rev. \textbf{D68}
(2003) 067301 [hep-th/0307239];

\bibitem{Wang} X.H.Meng and P.Wang, hep-ph/0311070;
\bibitem{WMAP} D.N.Spergel, et al., Astrophys.J.Suppl. \textbf{148} (2003) 1 [astro-ph/0302207]
; L.Page et al. astro-ph/0302220; M.Nolta, et al,
astro-ph/0305097; C.Bennett, et al, Astrophys.J.Suppl.
\textbf{148} (2003) 175 [astro-ph/0302209];
\bibitem{Linde} C.R.Contaldi, M.Peloso, L.Kofman and A.Linde, JCAP \textbf{0307} (2003)
002 [astro-ph/0303636]; J. M. Cline, P. Crotty, J. Lesgourgues,
JCAP \textbf{0309} (2003) 010 [astro-ph/0304558];
\bibitem{piao} Y.-S. Piao and Y.-Z. Zhang, astro-ph/0401231;
\end{thebibliography}
\end{document}